\documentclass[aps,prl,floatfix,showpacs,amsmath,amssymb,twocolumn,superscriptaddress]{revtex4-1}
\usepackage{graphicx}

\usepackage{setspace}
\usepackage{amsmath}
\usepackage{amssymb}

\begin{document}
\title{Granular Structure Determined by Terahertz Scattering}

\author{Philip Born}
\affiliation{Institut f\"ur Materialphysik im Weltraum,
Deutsches Zentrum f\"ur Luft- und Raumfahrt, 51170 K\"oln, Germany}
\author{Nick Rothbart}
\affiliation{Institut f\"ur Planetenforschung,
Deutsches Zentrum f\"ur Luft- und Raumfahrt, 51170 K\"oln, Germany}
\affiliation{Technische Universit\"at Berlin, Institut f\"ur
Optik und Atomare Physik, Stra\ss e des 17. Juni 135, 10623 Berlin,
Germany}
\author{Matthias Sperl}
\affiliation{Institut f\"ur Materialphysik im Weltraum,
Deutsches Zentrum f\"ur Luft- und Raumfahrt, 51170 K\"oln, Germany}
\author{Heinz-Wilhelm H\"ubers}
\affiliation{Institut f\"ur Planetenforschung,
Deutsches Zentrum f\"ur Luft- und Raumfahrt, 51170 K\"oln, Germany}
\affiliation{Technische Universit\"at Berlin, Institut f\"ur
Optik und Atomare Physik, Stra\ss e des 17. Juni 135, 10623 Berlin,
Germany}

\date{\today}
\begin{abstract}

Light-scattering in the terahertz region is demonstrated for granular matter. 
A quantum-cascade laser is used in a benchtop setup to determine the 
angle-dependent scattering of spherical grains as well as coffee powder and 
sugar grains. For the interpretation of the form factors for the scattering 
from single particles one has to go beyond the usual Rayleigh-Gans-Debye 
theory and apply calculations within Mie theory. In addition to single 
scattering also collective correlations can be identified and extracted as a 
static structure factor.

\end{abstract}



\maketitle

The statistical physics of granular matter has seen rapid progress in recent 
years: In addition to macroscopic measurements, experiments with direct 
imaging in two dimensions \cite{Majmudar2007} or tomography in three 
dimensions \cite{Jerkins2008} allow for analyses on the scale of individual 
particles. Many computer-simulation studies provide testable predictions for 
particle packings or reveal intriguing structural anomalies \cite{Donev2005}. 
Below the length scales of typical granular particles lies the realm of 
colloidal suspensions where light scattering has emerged as a reliable 
technique for the investigation of such structural properties and phase 
transitions \cite{Pusey1986}. Scattering offers the principle advantages of 
(a) being able to monitor the time evolution of the measured system, (b) good 
resolution of both larger and smaller length scales with reliable statistics, 
and (c) the in-situ measurement of three-dimensional samples. Hence, the use 
of scattering techniques is desirable also for granular matter.

With practical laser 
sources now emerging in the terhertz region, light scattering with 
wavelengths between 30~$\mu$m and 1~mm becomes possible \cite{Hubers2012}. 
As these wavelengths match typical particle sizes in granular media 
\cite{Rhodes1991,Duran1999}, but also sizes of cells and unicellular 
organisms \cite{Campbell2003}, suitable setups as well as methods for 
interpretation shall unlock the potential for terahertz light scattering 
in these areas. In the following, we present a bench-top light-scattering 
configuration using terahertz radiation, and demonstrate it for the 
investigation of granular particles between 80~$\mu$m and 1~mm in 
diameter. By suitable interpretation of the scattering intensities, we 
determine the particle sizes and for high densities identify the 
correlations among the particle positions in a static structure factor. 
The results demonstrate the possibility to investigate structures on 
previously unattainable length scales by scattering techniques in the 
terhertz region.

THz extinction spectra obtained from samples 
comprising granular media have already indicated sensitivity of the 
extinction to particle size and packing density 
\cite{Zurk2007,Bandyopadhyay2007,Kaushik2012a,Kaushik2012b}. However, such 
measurements are superpositions of spectroscopic features of the sampled 
materials as well as scattering effects, and therefore difficult to 
interpret. Measurement of the angular dependence of scattered intensities 
at a single wavelength leads to sensitivity to geometric features of the 
sample alone.

\begin{figure}[ht]
\centering
\includegraphics[width=0.45\textwidth]{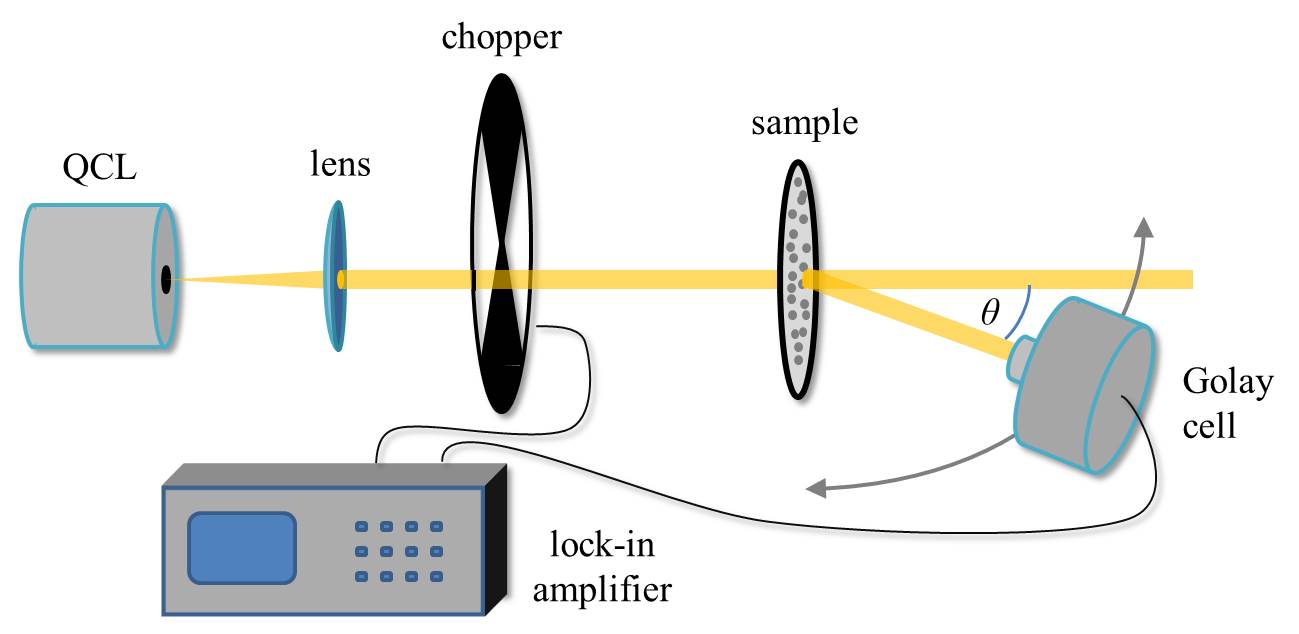}
\caption{\label{fig:setup}Setup for THz scattering from granular matter. A 
beam from a quantum cascade laser (QCL) operating at 3.4~THz is chopped in 
a lock-in configuration to suppress thermal background radiation. The 
incoming beam is scattered from a quasi-two dimensional granular sample at 
variable angle $\theta$. A Golay-cell is used as a detector.}
\end{figure}

The development of terahartz quantum cascade lasers (QCLs) 
\cite{Kohler2002,Hubers2010} make experiments feasible that are in close 
analogy to static light scattering (SLS) setups which apply visible light. 
In SLS, lasers provide collimated, monochromatic, high intensity 
radiation. Common solvents like water, organic liquids and air are 
transparent, and background light is easily shielded by lightproof 
containers. Using THz radiation imposes several constraints not present in 
experiments with visible light: Many media including ambient air are 
highly absorbing over large regions of the THz spectrum, and strong 
background radiation from thermal emission at room temperature has to be 
faced \cite{Hubers2012}. To overcome these constraints we use (1) a 
transmission window of air at 3.4~THz, (2) thin, effectively 
two-dimensional samples and (3) lock-in detection of the signal. 
Consequently, we obtained a versatile bench-top experiment for 
angle-resolved scattering experiments (see Fig.~\ref{fig:setup}). The 
capabilities of the setup are demonstrated in the following to 
characterize spherical particles with diameters from 80~$\mu$m to 
1000~$\mu$m as well as technical grade coffee powder and sugar grains.

First, polystyrene and polyethylene spheres are used with nominal diameters of 
80~$\mu$m ('PS 80'), 250~$\mu$m ('PS 250'), 500~$\mu$m ('PS 500'), and 
1000~$\mu$m ('PE 1000'). From microscopy images we obtained the particle 
radius $a$, the standard deviation $\sigma$ and the polydispersity 
$\text{PD}=\sigma/a\cdot100$ of the PS and PE particles. The size ratios 
$x=2\pi\cdot a/\lambda$ and the phase shifts $\rho=2x\cdot|m-1|$ were 
calculated using $a$ and literature values for the complex refractive indices 
relative to air of $m_{\text{\small PS}}=1.59-i\cdot0.002$ and 
$m_{\text{\small PE}}=1.54-i\cdot0.0014$ \cite{Cunningham2011}. The number of 
illuminated particles followed from the cross-section of the laser beam of 
3~mm$^{2}$ and the packing fraction of 0.55 determined from microscopy images. 
The scattering vector is determined by the scattering angle $\theta$ and the 
wavelength in air $\lambda$, $q=4\pi / \lambda \cdot \sin(\theta/2)$. From the 
packing density and the beam cross section, the number of illuminated 
particles was estimated (see Supplemental Material for description and 
determination of parameters). Table~\ref{tab:samples} summarizes the 
characteristic parameters of the samples.

\begin{table}[ht]
\centering
\small
  \caption{Summary of the sample characterization. $a$ is the particle 
radius and PD is the polydispersity of the particles as measured by light 
microscopy, $x$ the size ratio, $\rho$ the phase shift, and $n_{ill}$ the 
number of particles illuminated by the primary beam.}
  \label{tab:samples}
  \begin{tabular}{lccccc}
    \hline \hline
    name & a [$\mu$m] & PD [\%] & $x$ & $\rho$ & $n_{ill}$  \\
    \hline
    PS 80 & 40.3 & 9.2 & 3.028 & 3.57 & $\approx$ 250 \\
    PS 250 & 109.5 & 5.1 & 8.327 & 9.83 & $\approx$ 40 \\
    PS 500 & 286.1 & 4.6 & 21.65 & 25.55 & $\approx$ 5 \\
    PE 1000 & 500 & n.d. & 37.85 & 40.88 & $\approx$ 2 \\
    \hline \hline
  \end{tabular}
\end{table}

A custom-made QCL, obtained from the Paul-Drude-Institut f\"ur 
Festk\"orperelektronik, Berlin, provided linearly polarized THz radiation 
with an average wavelength in air of 87.2~$\mu$m. We measured in 
1.5$^{\circ}$ steps from -20$^{\circ}$ to 100$^{\circ}$ around the sample 
with an angular resolution of 2.6$^{\circ}$ using a Golay cell. A thin PE 
foil supported the particles with an angle of 20$^{\circ}$ to the primary 
beam to allow asymmetric measurements up to scattering angles of 
100$^{\circ}$. Background intensity measurements with empty PE foil showed 
scattering up to 20$^{\circ}$, the full-width at half-maximum was 
determined to 14.1$^{\circ}$. The obtained scattering spectra from 
particles were corrected and normalized for background and drift in 
detector response and QCL power.

\begin{figure}[htb]
\centering
\includegraphics[width=0.45\textwidth]{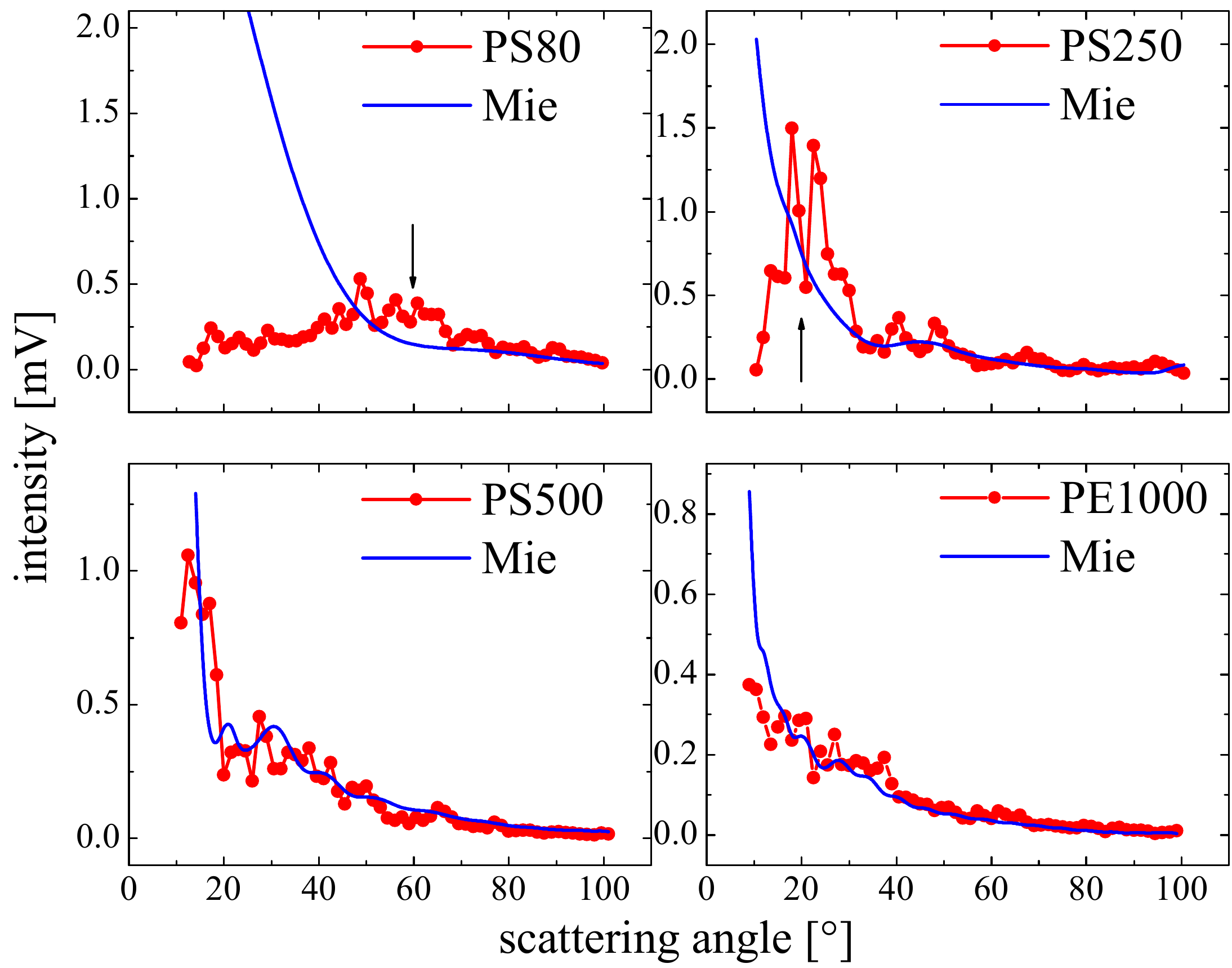}
\caption{\label{fig:mieint}Angle-resolved THz scattering intensities for 
spherical particles. Mie-theory predictions for scattering from single 
spheres fit the experimental background-corrected scattered intensities of 
polystyrene (PS) and polyethylene (PE) samples. Deviations at small angles 
in PS 80 and PS 250 indicate scattering from particles with correlated 
positions (see text). Arrows for PS 80 and PS 250 show the scattering 
angle below which length scales larger than the particle diameter are 
probed and deviations are expected.}
\end{figure}

The intensity of the radiation scattered by the particles exhibits variations 
with angle and particle size. Figure~\ref{fig:mieint} shows the measured 
intensities for scattering angles from $\theta_{\text{min}} = 12^{\circ}$ up 
to $\theta_{\text{max}} = 100^{\circ}$. In a first step to interpret the 
measured variations, we compare the experimental data with theoretical 
predictions for scattering from single spheres. From this we shall obtain 
quantitative information on the particle size and discuss the systematic 
deviations from theory in terms of correlated particle positions. The phase 
shifts $\rho$ produced by the particles (see Tab.~\ref{tab:samples}) exceeds 
the assumption underlying Rayleigh-Gans-Debye (RGD) theory by far: $\rho<1$ is 
commonly used in SLS-measurements \cite{Egelhaaf2006}, while phase shifts and 
size ratios larger than 1 indicate a much better applicability of the 
approximative theories of anomalous diffraction by van de Hulst (vdH) 
\cite{Hulst1981}. We calculated scattered angle-dependent intensities 
according to RGD and vdH as well as Mie theory. Fitting of the data was 
performed using Matlab codes (www.mathworks.com). Scattering amplitudes for 
anomalous diffraction were calculated using the first three terms of the vdH 
series \cite{Hulst1981}. The Mie solution to the scattering problem was 
calculated using $x+4\cdot x^{1/3}+2$ scattering coefficients in the 
Matlab-implementation by C.~M\"atzler \cite{Matzler2002} of the Bohren-Huffman 
code \cite{BH1983} following the Wiscombe critaeria \cite{Wiscombe1980}, cf. 
details in the Supplementary Material.
The scattering amplitude as calculated from RGD theory depends only on the 
particle radius and the scattering angle. Scattering amplitudes for 
anomalous diffraction additionally take into account the phase shift and 
the full complex refractive index of the particles \cite{Hulst1981}. The 
polarization of the THz radiation perpendicular to the scattering plane is 
taken into account in the Mie calculations. Additionally, the divergence 
of the primary beam was taken into account by a moving average over 
scattering angles of 15$^{\circ}$ (see Supplemental Material for details of 
the used models).

\begin{figure}[ht]
\centering
\includegraphics[width=0.3\textwidth]{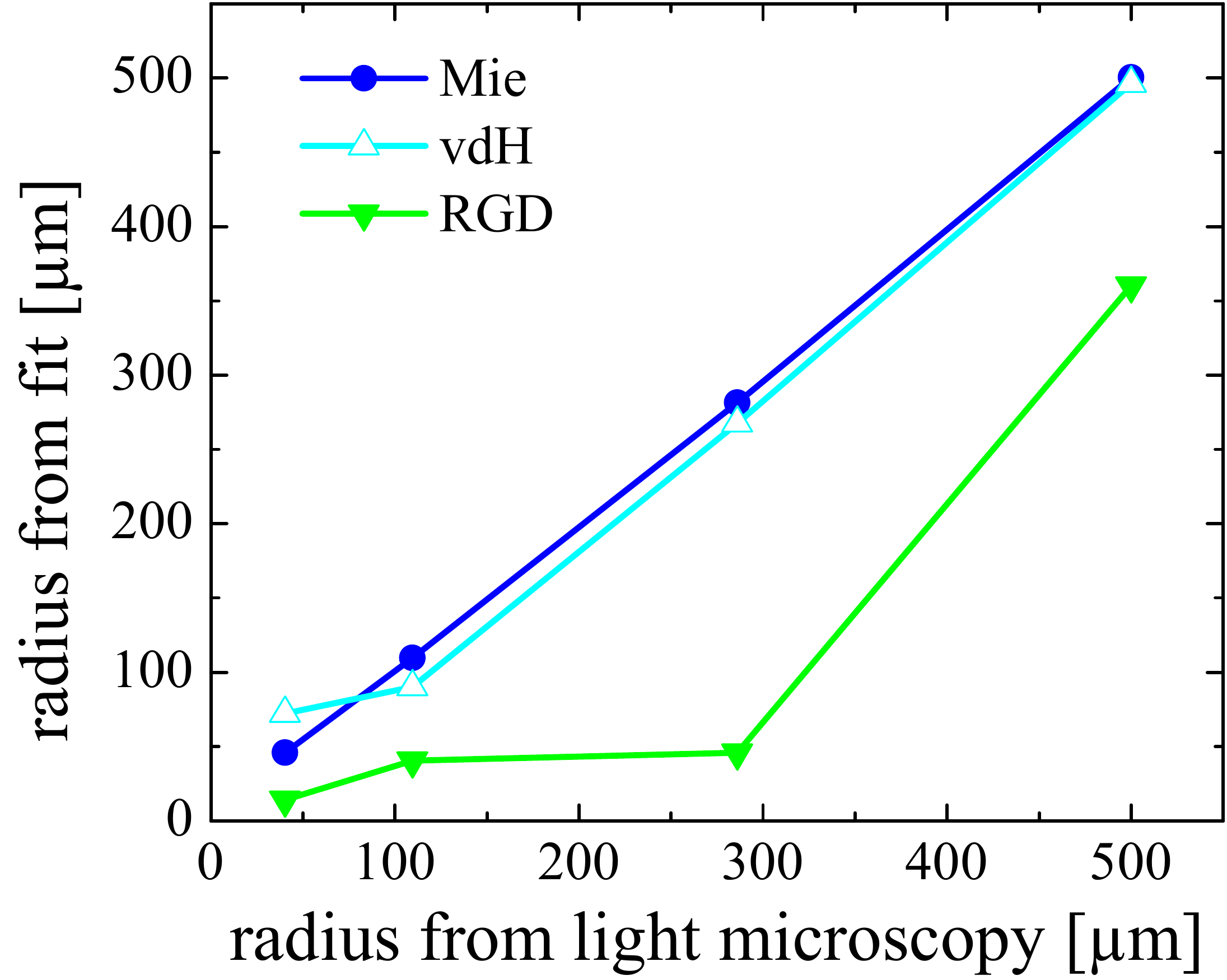}
\caption{\label{fig:sizes}Determination of the particle radius by 
scattering. Particle radii measured by least-square fits of Mie-theory 
(Mie) and predictions by the approximative theories of van de Hulst (vdH) 
and Rayleigh-Gans-Debye (RGD) are compared to microscopy measurements. The 
results from the Mie-theory are nearly indistinguishable from a bisecting 
line.}
\end{figure}

The agreement among the predictions of the Mie theory and the measured 
intensities can be seen in Fig.~\ref{fig:mieint}. Fitting of the 
experimental scattering data with the theoretical models allows 
determining the particle radius. Excellent agreement can be seen in 
Fig.~\ref{fig:sizes} between the particle radius as obtained from 
scattering and Mie Theory as compared to light microscopy. The vdH theory 
is nearly in as good agreement with the microscopy data as the Mie theory, 
while RGD theory systematically underestimates particle sizes.

All applied scattering theories are strictly valid only for scattering 
from individual spheres with uncorrelated positions. For the smaller 
spheres, PS 80 and PS 250, deviations by correlations among the particle 
positions can be expected. The most pronounced effect of interference of 
light scattered from particles with correlated positions is the 
suppression of scattering at small scattering angles \cite{Brown1996}. 
This can be observed in the scattering spectra of PS 80 and PS 250.
Arrows in Fig.~\ref{fig:mieint} indicate where 
the respective scattering angles correspond to a length scale of the 
particle diameter, $\sin(\theta^*/2) = \lambda/(4a)$. Smaller scattering 
angles, $\theta \lesssim \theta^*$, hence describe length scales larger 
than the particle diameters. The Mie prediction for single-sphere 
scattering alone overestimates scattered intensities at these angles.
For our thin samples this suppression can be described by the structure factor 
$S(q)$, the Fourier-transform of the point pattern representing the particle 
positions \cite{Brown1996}. 
The scattered intensity $I_{\text{sca}}$ becomes 
the product of the single-sphere scattering amplitude $P(q)$ and the structure 
factor $S(q)$: $I_{\text{sca}}(q) \propto P(q)\cdot S(q)$, where $q$ is the 
modulus of the scattering vector.

\begin{figure}[ht]
\centering
\includegraphics[width=0.45\textwidth]{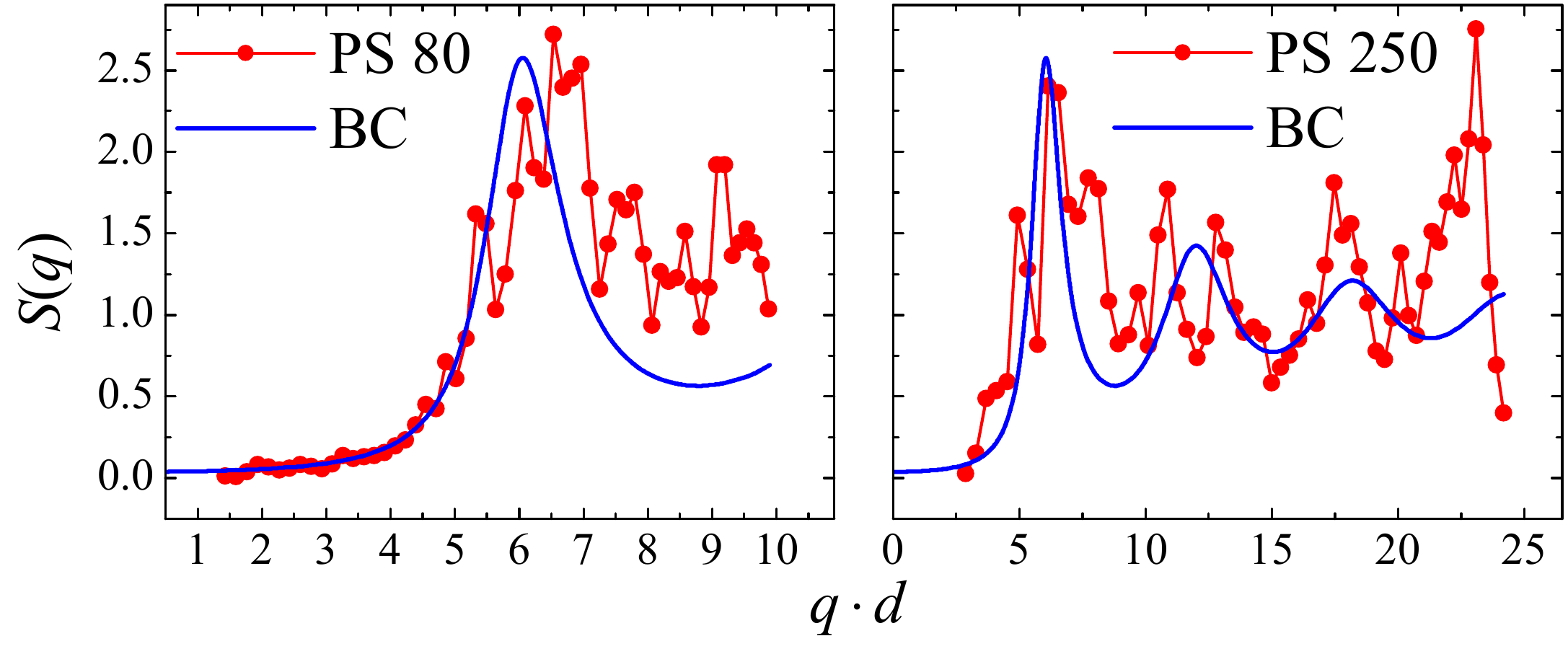}
\caption{\label{fig:sq}Experimental structure factors from THz scattering.
Structure factors of the samples are determined by $S(q) = I_\text{sca}(q) 
/ P_\text{Mie}(q)$ and shown together with the Baus-Colot (BC) prediction 
for hard-disc fluids with a packing fraction of 0.6. The scattering vector 
is scaled with the particle diameter.}
\end{figure}

We divide the scattered intensity by the result of the Mie prediction for 
scattering from a single sphere $P_\text{Mie}(q)$ to get an estimate of 
the structure factor. Figure~\ref{fig:sq} shows the result for this 
calculation along with the prediction of the Baus-Colot theory for the 
structure factor of two-dimensional hard-disc liquids with an area 
fraction of 0.6 \cite{Baus1986}. The position of the first maximum and the 
asymptotic behavior of the structure factor for small $q$-values agree 
well with the analytical prediction for hard-disc fluids. Additional 
fluctuations in the experimental structure factor arises from the limited 
number of configurations probed in the static experiments. To compare the 
measured structure factor with the simulation, we estimate from Fig.~2 in 
\cite{Donev2005} that the predicted anomaly $S_q\propto q$ is found for wave 
vectors $qd\lesssim 3$ in the simulation. As seen in the left panel of 
Fig.~\ref{fig:sq}, in the experiment the wave vectors for the 80$\mu$m 
particles start around $qd\gtrsim 1$ allowing for enough overlap to give a 
first estimate about $S_q$ in the indicated regime. The Baus-Colot fit would
indicate within the experimental resolution both the absence of a linear 
regime in $S_q$ as well as the existence of a finite rather than a vanishing 
intercept for $qd\rightarrow 0$ at the given particle density. Given the low
intensities in that respective wave-vector regime, caution is appropriate not
to overstate that finding. While our 
results can not yet be compared with the simulation at close packing, the 
granular structure factors away from the transition point are consistent with 
the simulation \cite{Donev2005} and show the applicability of THz scattering
in the relevant regime of wave vectors.

\begin{figure}[ht]
\centering
\includegraphics[width=0.45\textwidth]{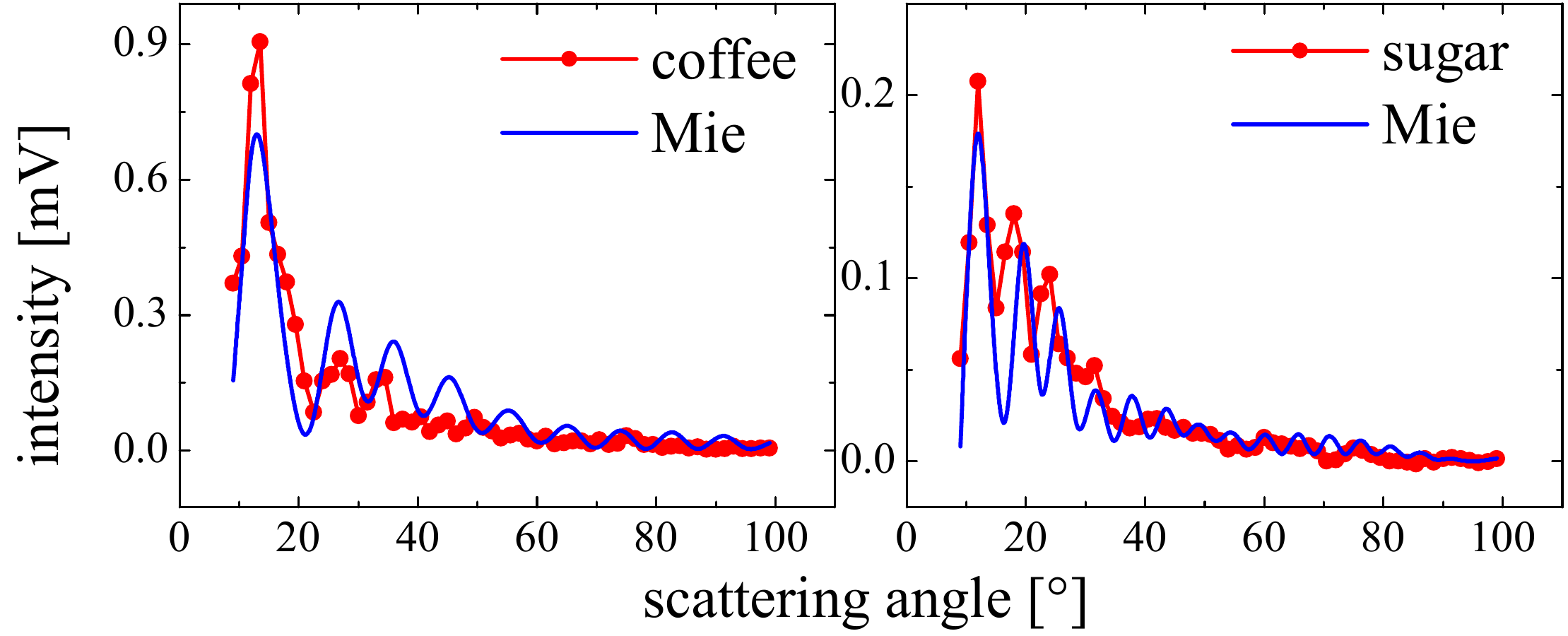}
\caption{\label{fig:cofsug}Background-corrected scattered intensities of 
coffee powder (left) and sugar grains (right). Curves indicate
least-square fits of Mie-theory calculations used to determine the 
particle sizes.
}
\end{figure}

To prove the applicability to technical grade particles, we demonstrate 
scattering from coffee powder and sugar grains in Fig.~\ref{fig:cofsug}. 
We show the background-corrected scattered intensities along with the fit 
of Mie-scattering curves. The refractive index of polystyrene was used in 
the fitting procedure. The imaginary part of the refractive index had to 
be increased to improve fit quality. From the Mie prediction one can fit 
the particle radii of 300~$\mu$m for coffee powder and 600~$\mu$m for 
sugar grains, respectively. This is within expectation for the 
millimeter-sized sugar grains, and may indicate a sensitivity to 
agglomerates of the 300~$\mu$m large coffee grains.

In summary, we demonstrated a versatile bench-top setup for static THz 
radiation scattering. With this setup, size-dependent angle-resolved 
scattered intensities can be measured. By using Mie theory or the 
van~de~Hulst approximations, THz radiation scattering can be used to 
determine particle sizes. For quasi two-dimensional particle packings, 
also static structure factors can be measured, which indicate the 
correlations among the particle positions. Concludingly, we have shown 
that by closing the gap between the infrared and the microwave part of the 
electromagnetic spectrum, a window opens for scattering methods to 
determine structural properties of particles with submillimeter sizes.

We thank H.~T.~Grahn, R.~Hey, L.~Schrottke, and M.~Wienold from the 
Paul-Drude-Institut f\"ur Festk\"orperelektronik for providing the QCL. N.~R. 
acknowledges support from the Helmholtz-Research School on Security 
Technologies. P.~B. thanks P.~Kuhn for discussions and A.~Meyer for continued 
support.

\newpage
\normalsize
\section*{Supplemental Material}

\begin{figure}[htb]
\centering
  \includegraphics[width=0.45\textwidth]{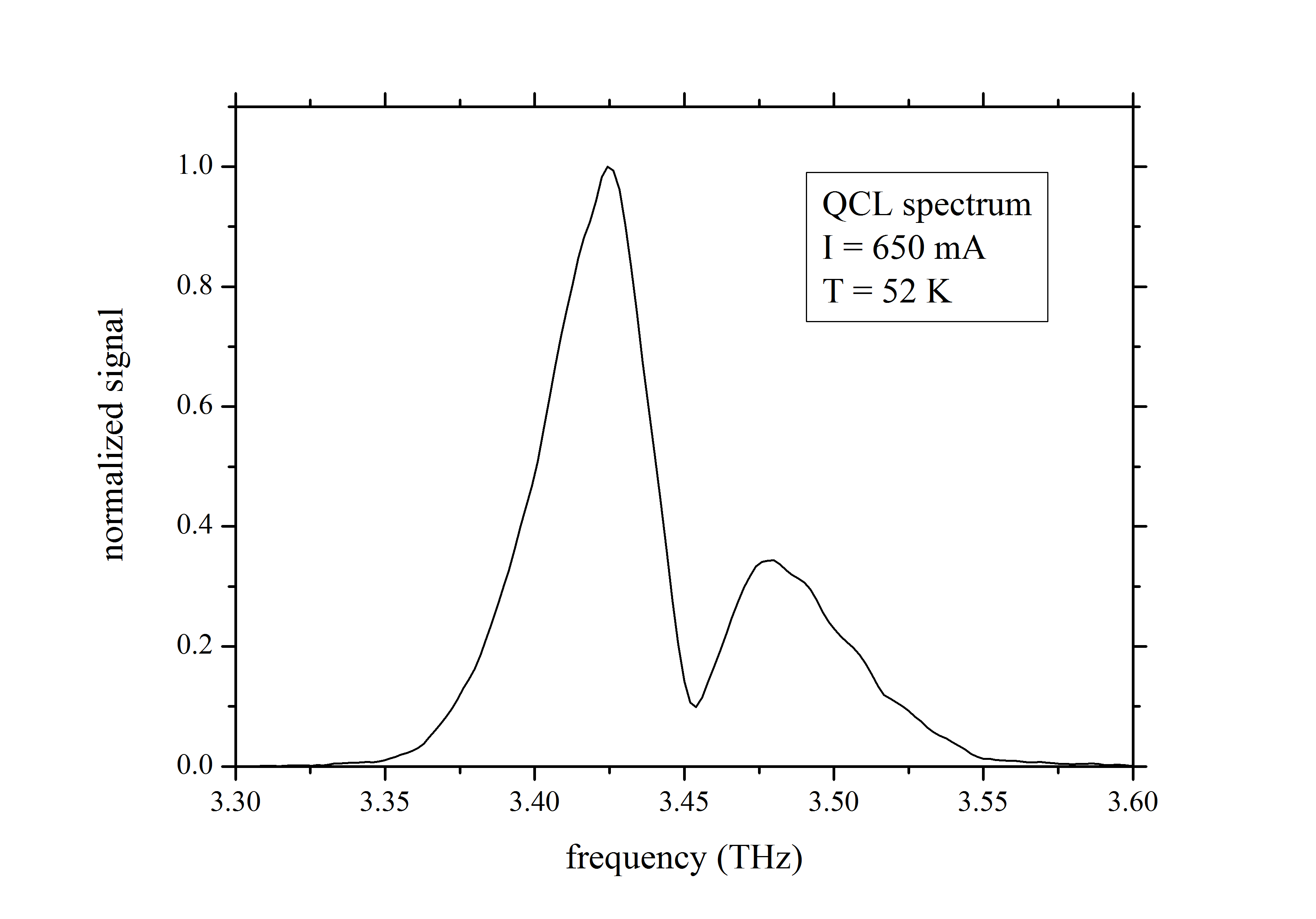}
  \caption{Spectrum of the radiation emitted by the quantum 
cascade laser at 52~K. Two modes are observable, the peak heights are 
used to determine the average wavelength of the laser in air.}
  \label{fig:qclspec}
\end{figure}

\begin{figure}[htb]
\centering
  \includegraphics[width=0.45\textwidth]{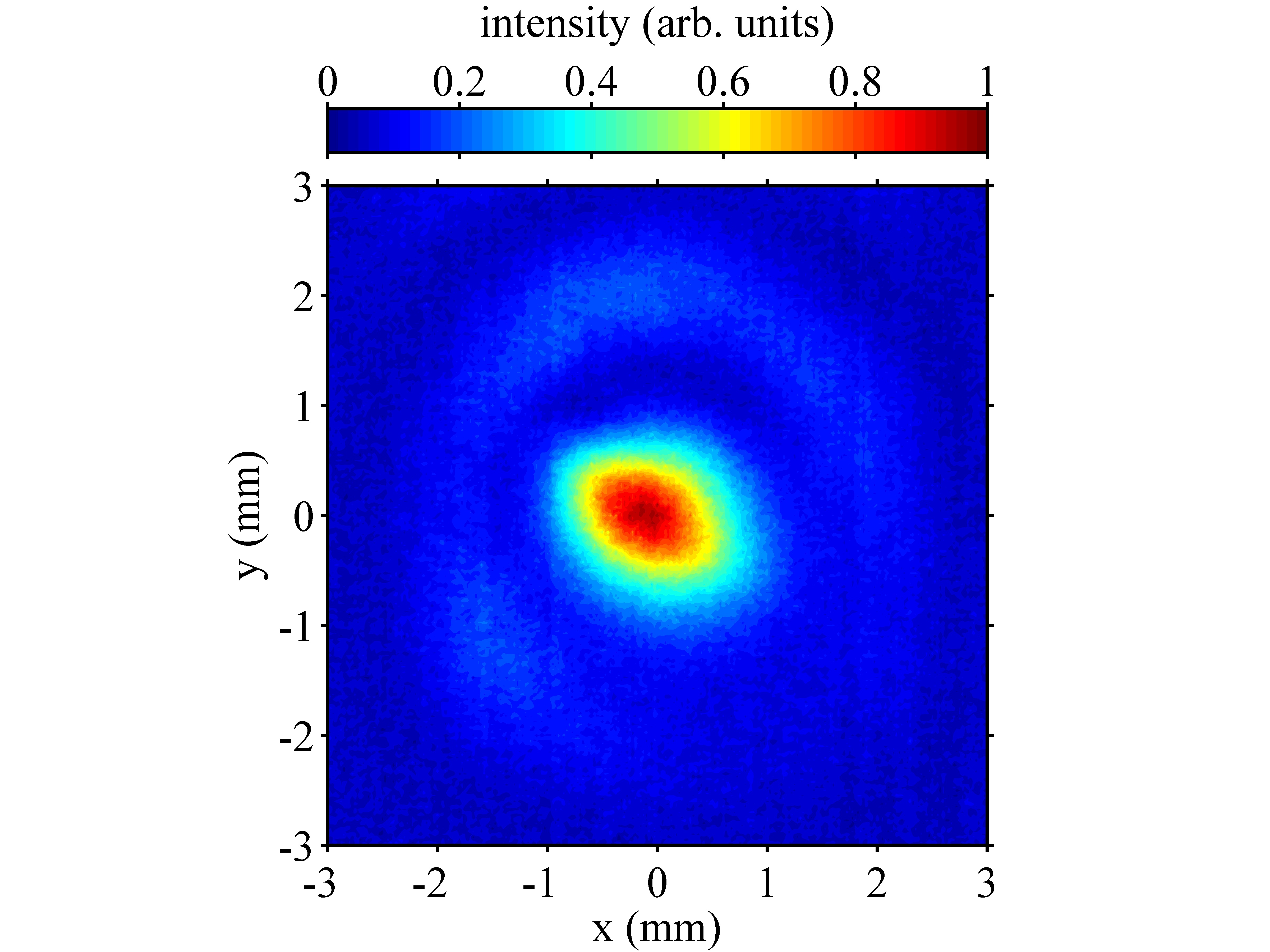}
  \caption{Intensity profile of the laser beam. At the 
position of the sample the beam has a nearly Gaussian shape.}
  \label{fig:cross}
\end{figure}

\begin{figure}[htb]
\centering
  \includegraphics[width=0.45\textwidth]{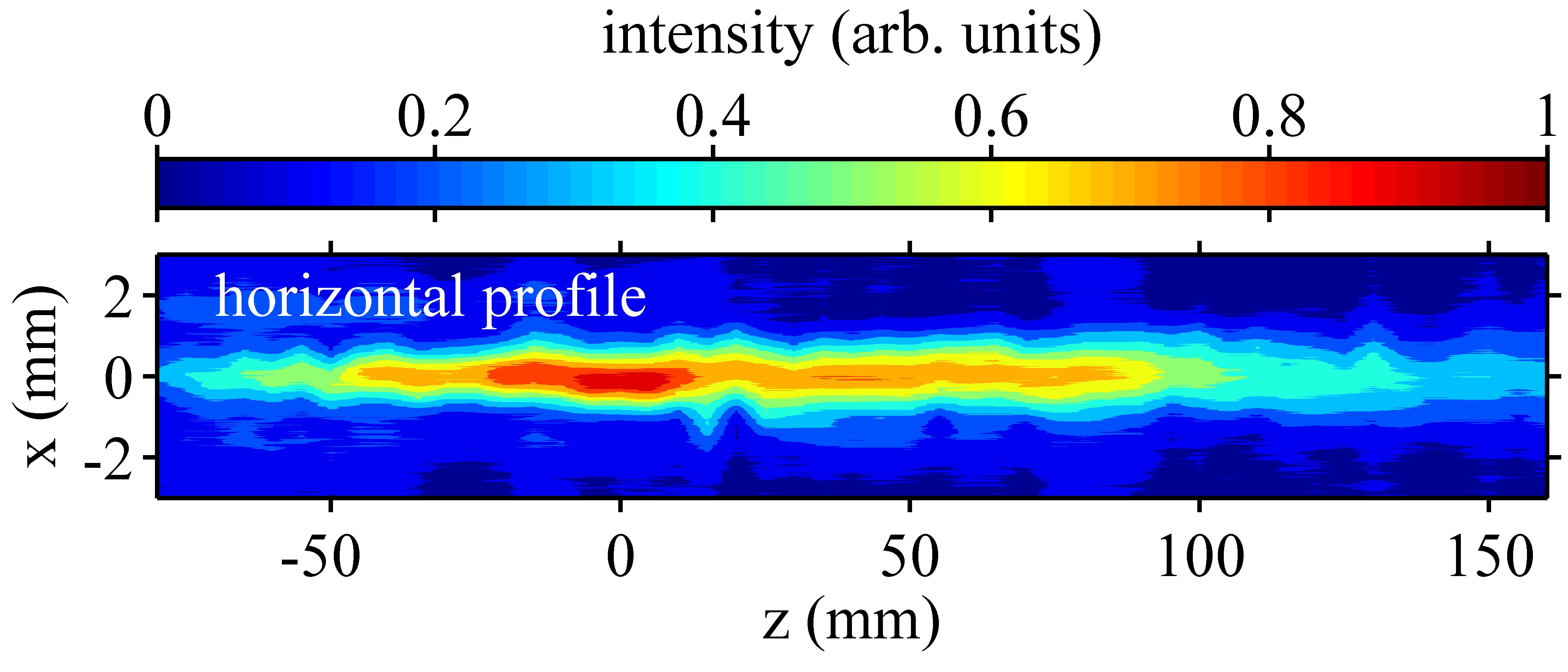}
  \includegraphics[width=0.45\textwidth]{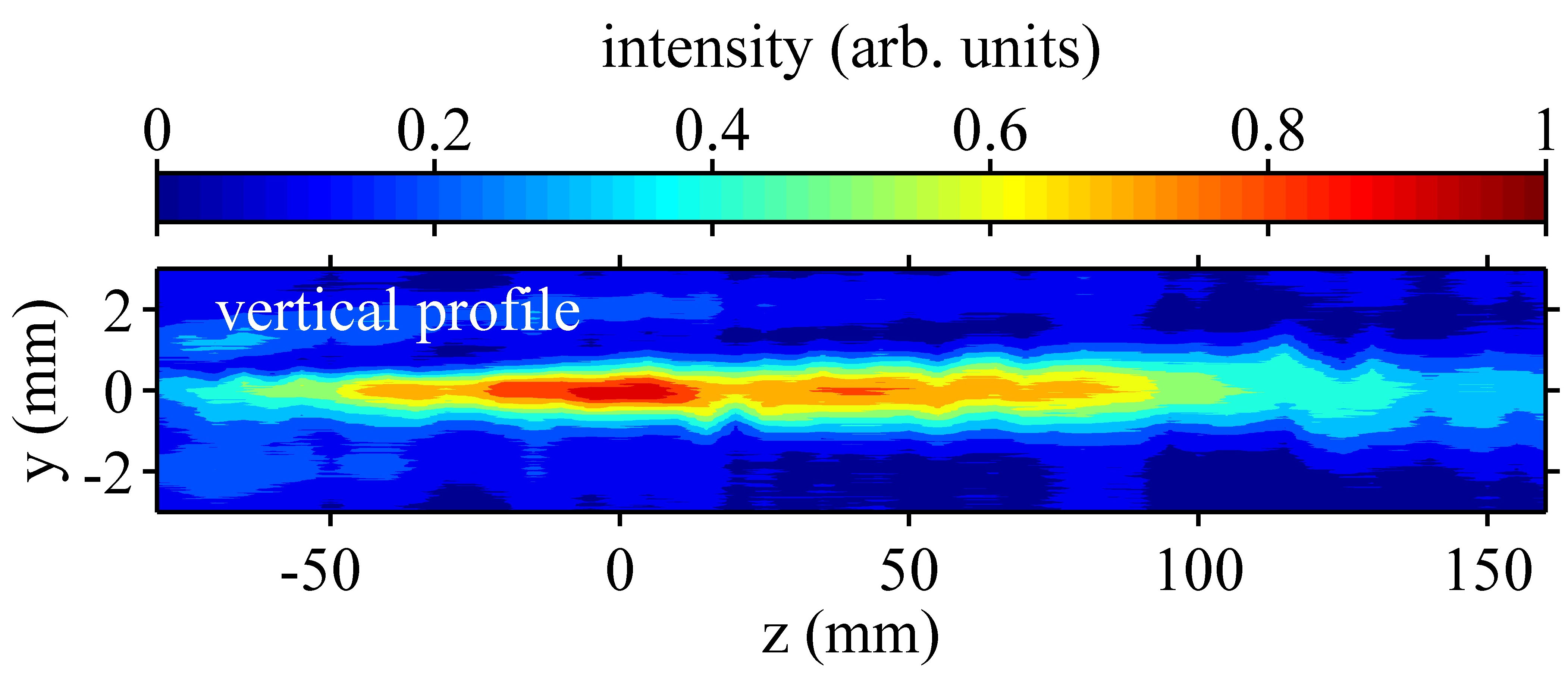}
  \caption{Profile of the laser beam in the scattering plane (top) and 
in the plane perpendicular to the scattering plane (bottom).
The sample is located at $z = 0$.}
  \label{fig:beamx}
\end{figure}

\paragraph{Setup} A custom-made quantum cascade laser (QCL), obtained 
from the Paul-Drude-Institut f\"ur Festk\"orperelektronik 
\cite{Hubers2010}, Berlin, provided the required radiation. At 52~K, the 
laser emitted approximatively 2.2~mW in two modes at 3.42~THz 
(87.55~$\mu$m) and 3.48~THz (86.15~$\mu$m) linearly polarized 
perpendicular to the scattering plane. An emission spectrum can be found 
in Fig.~\ref{fig:qclspec}. From the peak heights an 
average frequency of 3.44~THz (87.2~$\mu$m) was determined. A plano-convex 
TPX lens was adjusted 90~mm from the laser in the light path to minimize 
beam divergence. Measurements with a modified mid-IR microbolometer camera 
(Infratec Variocam) gave a beam cross-section of 3~mm$^{2}$ at the sample 
position (see Figs.~\ref{fig:cross} and \ref{fig:beamx}).

A goniometer (PI M-660.55 rotation stage) was installed 600~mm from the 
lens for angle-resolved measurements. A Golay-cell (Tydex GC-1T) measured 
the light intensity on an extension arm 55~mm from the center of the 
goniometer. That led to a total free path length in air of 745 mm. A 
2.5~mm wide circular diaphragm in front of the cell entrance window 
ensured an angular resolution of 2.6$^{\circ}$.

The application of a Golay-cell as detector and the strong thermal 
background made chopping the light signal and lock-in detection necessary. 
We chopped the light with 10~Hz and detected with 2~mV lock-in sensitivity 
(Stanford Research SR 850 amplifier). This led to saturation of the 
detector signal in the primary beam but ensured sensitivity to weak 
scattering signals.

\begin{figure}[htb]
  \includegraphics[width=0.45\textwidth]{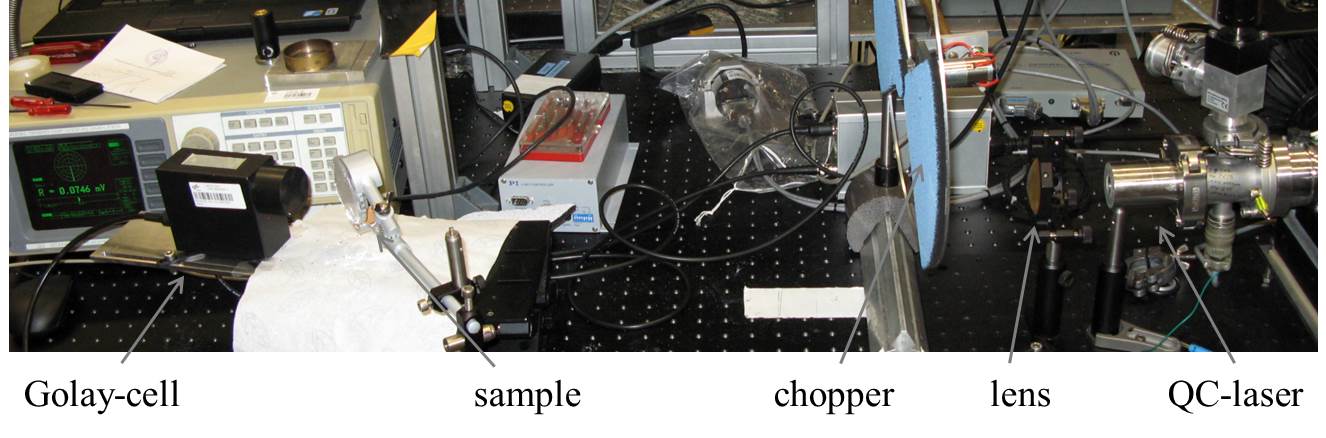}
  \caption{Photograph of the benchtop experimental setup.}
  \label{fig:photosetup}
\end{figure}

A very thin sample holder was necessary to minimize absorption losses. 
10~$\mu$m thin polyethylene (PE) foil was tautened over a metal ring with 
7.5~cm diameter. By pressing and removing adhesive tape onto and from the 
foil a thin residual adhesive film was created. The sample holder took an 
angle of 20$^{\circ}$ with respect to the laser beam to allow asymmetric 
measurements up to high scattering angles of 100$^{\circ}$ without 
shadowing by the metal ring. A photograph of the assembled setup is 
available in Fig.~\ref{fig:photosetup}.

\begin{figure}[ht]
\centering
\includegraphics[width=0.45\textwidth]{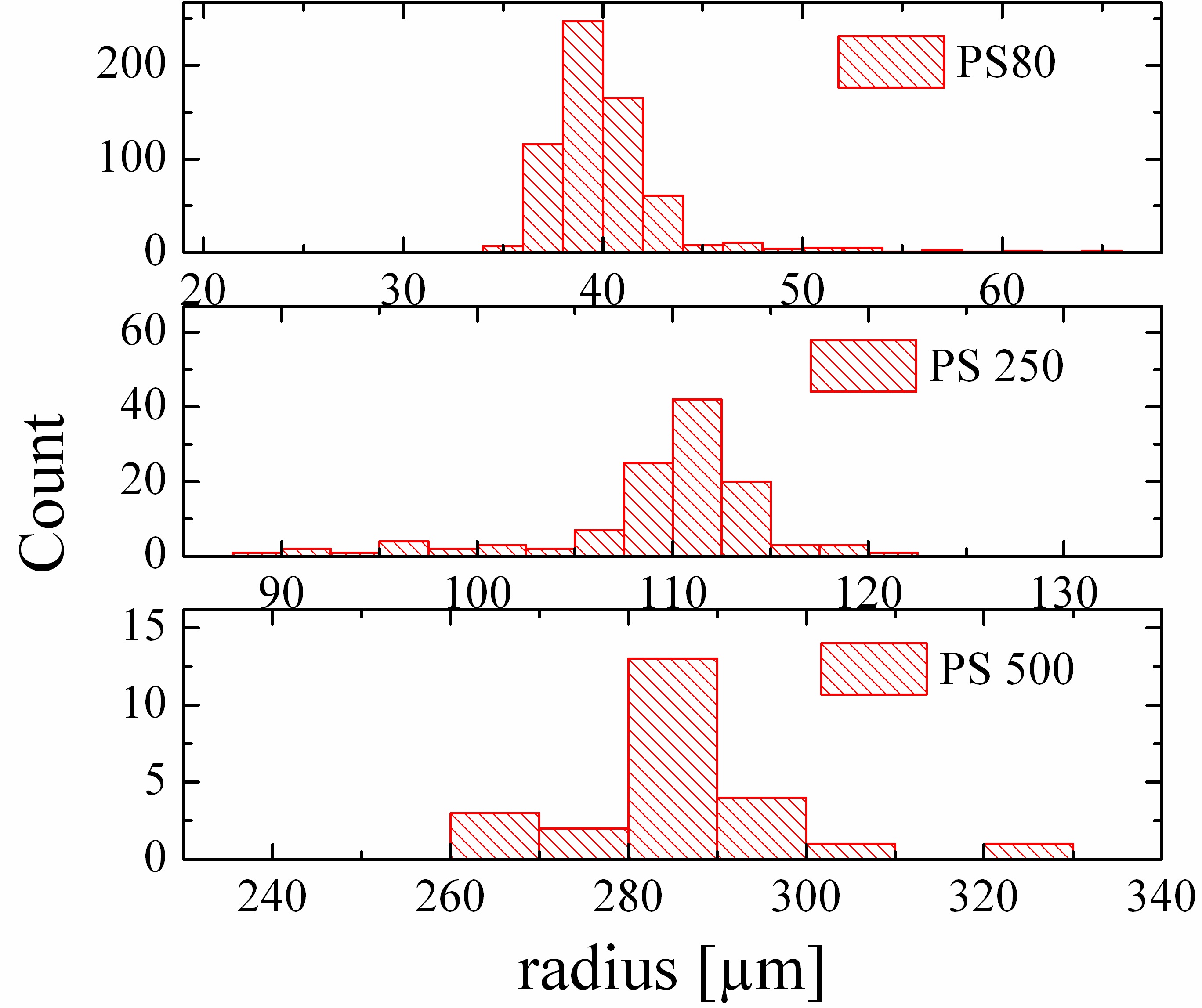}
\caption{\label{fig:part}Size distribution of the polystyrene 
samples as measured with light microscopy.}
\end{figure}

\paragraph{Samples} Polystyrene (Dynoseeds, Microbeads AS) and 
polyethylene spheres (Sphero, Rieter GmbH) were used as purchased. Light 
microscopy images were taken of the polystyrene samples and evaluated 
using ImageJ (http://rsbweb.nih.gov/ij/). The mean particle size and the 
size distribution was determined from the light microscopy images 
(Fig.~\ref{fig:part}). 

\begin{figure}[htb]
\centering
  \includegraphics[width=0.4\textwidth]{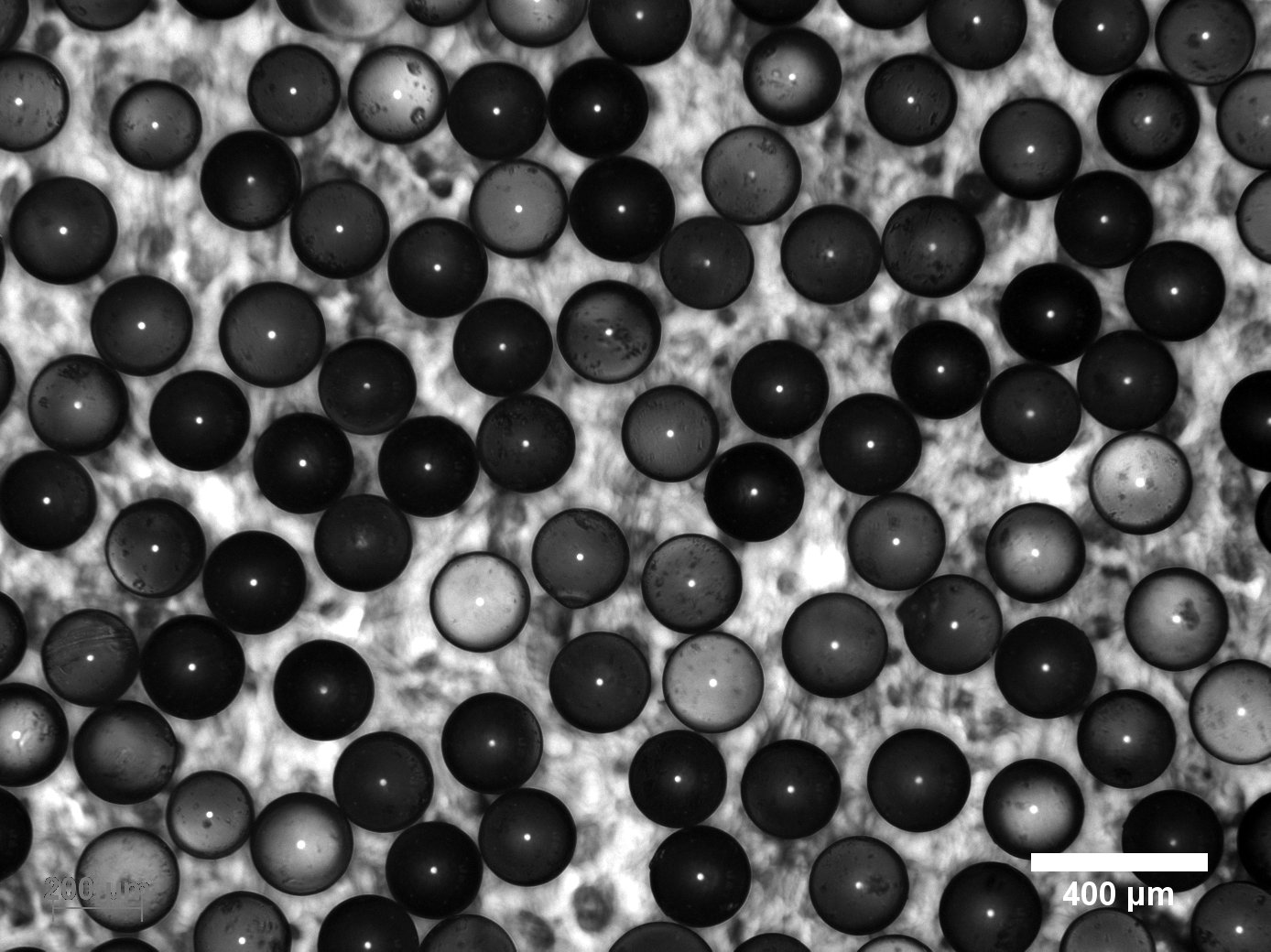}
\caption{\label{fig:ps250}Representative image of the 
deposited granular particle samples, here PS 250.}
\end{figure}

The polydispersity PD was defined by the standard deviation $\sigma$ of the 
mean radius $a$ and the mean radius $a$, $\text{PD}=\sigma/a\cdot100$. The 
particles were spread on the adhesive film on the PE foil; subsequently 
unfixed particles were blown off, yielding a packing density of deposited 
$\phi\approx~55\%$ (see Fig.~\ref{fig:ps250} for a microscopy image of the 
deposited particles). For calculation of $x$ and $\rho$ literature values for 
the complex refractive indices of PS and PE were used, 
$m_{PS}$=1.59-$i\cdot0.002$ and $m_{PE}$=1.54-$i\cdot0.0014$ 
\cite{Cunningham2011}

\begin{figure}[ht]
\centering
  \includegraphics[width=0.35\textwidth]{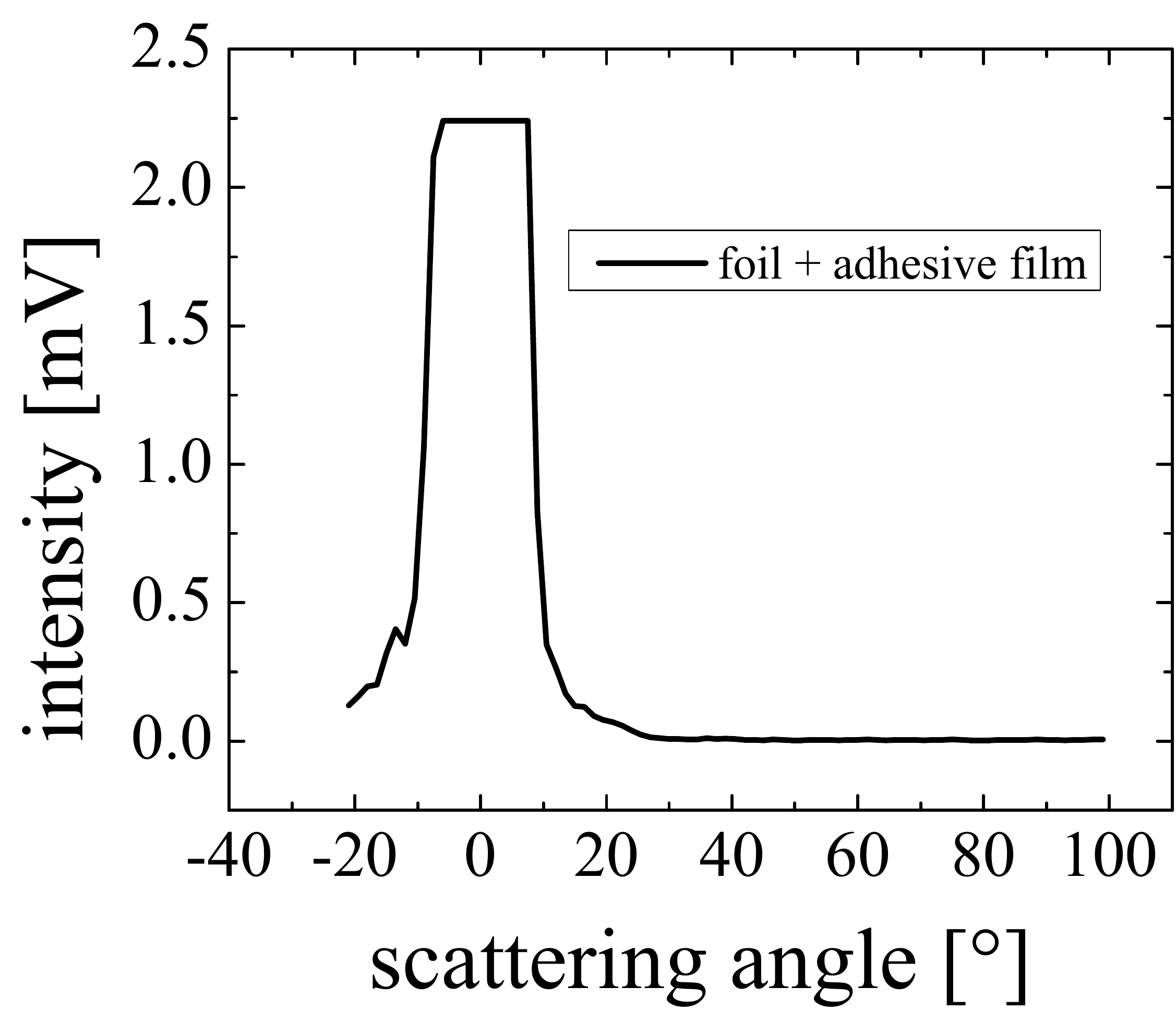}
\caption{\label{fig:bgint}Scattering spectrum of the laser 
beam after it passed through the PE foil. The signal in the range from 
-10$^{\circ}$ to 10$^{\circ}$ is saturated, a Gaussian fit yields a 
full-width half maximum of 14.1$^{\circ}$, above 20$^{\circ}$ no scattered 
intensity is detectable.}
\end{figure}

\paragraph{Size-dependent angle-resolved scattering} Measurements were 
performed from -20$^{\circ}$ to 100$^{\circ}$ in 1.5$^{\circ}$-steps. The 
intensity scattered by the PE foil and the adhesive film was measured prior to 
particle measurements. The beam exhibited scattering up to 20$^{\circ}$, the 
full-width half-maximum was determined to 14.1$^{\circ}$. Above 20$^{\circ}$ 
virtually no scattered intensity was detected (see Fig.~\ref{fig:bgint}). The 
scattering spectra from the particles were corrected and normalized for the 
background from the empty sample holder and the drift in detector response and 
QCL power.

\paragraph{Fitting} Fitting of the data was performed using Matlab 
codes 
and the least-square fitting routines provided by the optimization toolbox 
for Matlab (www.mathworks.com). A scaling factor and the particle radius 
was left as a free fit parameter. Three models for single sphere 
scattering were tested: Rayleigh-Gans-Debye theory (RGD), van de Hulst's 
approximation (vdH), and the exact Mie-solution (Mie). RGD theory provides 
a form factor $P_\text{RGD}(q)$ for scattering from a single sphere 
\cite{Hulst1981}, which depends only on the particle radius $a$ and the 
scattering vector $q=4\pi / \lambda \cdot \sin(\theta/2)$:
\begin{equation}
P_\text{RGD}(q,a) = 
\left[\frac{3}{(qa)^3}(\sin(qa)-qa\cos(qa))\right]^{2}.
\label{eq:rgd}
\end{equation}
RGD theory can be expected to give meaningful results only for particles 
with $\rho\ll1$. The fit quality using RGD theory was generally poor, we 
obtained root-mean-square(rms)-deviations of 6.5~mV (23.7\%, averaged over 
all particle sizes).

Van de Hulst dropped the limitation of $\rho$ being small and obtained a 
series expansion for the scattering from individual spheres 
\cite{Hulst1981}, albeit still with the restriction of the refractive 
index $m$ being close to unity, $|m-1|<1$. The series converges with 
$1/\rho^{2}$; we used the first three terms of the series expansion. Using 
the next higher term in the series yielded only relative changes of the 
scattering amplitude on the order of $10^{-2}$. The scattering amplitude 
in the vdH approximation takes the full complex refractive index into 
account:
\begin{eqnarray}
P_{\text{vdH}}(m,a,\theta) &=& \left[\frac{\rho}{y^{2}}\left(\frac{\pi y}{2}
\right)^{1/2}J_{3/2}(y)\right]^{2} \nonumber\\ & & + 
\left[\frac{1}{z}J_{1}(z) + \frac{\rho}{y^{2}}\left(\frac{\pi y}{2} 
\right)^{1/2}Y_{3/2}(y) \right. \\ & & \left. + 
\frac{1}{\rho^{2}}J_{0}(z) + \ldots{} \right]^{2}, \nonumber
\label{eq:vdh}
\end{eqnarray}
where $z=x\cdot\theta$, $y^{2}=\rho^{2}+z^{2}$, and $J_{\alpha}(y)$ and 
$Y_{\alpha}(y)$ being Bessel functions of the first and the second kind, 
respectively. The fit quality using vdH theory was very good, we obtained 
rms deviations of 0.4~mV (7.7\%, averaged over all particle sizes).

To calculate the exact Mie-theory solution to the scattering problem the 
Matlab-implementation of the Bohren-Huffman code \cite{BH1983} by 
C.~M\"atzler was used \cite{Matzler2002}. Following the criteria given by 
Wiscombe, $x+4\cdot x^{1/3}+2$ scattering coefficients were calculated to 
obtain the scattering amplitude \cite{Wiscombe1980}. The polarization of 
the THz radiation perpendicular to the scattering plane is taken into 
account in the Mie-calculations. The fit quality using Mie theory was 
excellent, we obtained rms-values of 0.2~mV (4.7\%, averaged over all 
particle sizes).

The full-width half-maximum of 14.1$^{\circ}$ of the primary beam without 
particles was taken into account in the comparison of the experimental 
scattering data to the theoretical predictions by a moving average over 10 
data points or 15$^{\circ}$, respectively.

\paragraph{Structure factor calculation}

The structure factor represents the phase shift between the light 
scattered by the individual particles; it is thus equivalent to the 
Fourier transform of the pair distribution function $g(r)$, the 
probability of finding at a distance $r$ from a first particle center a 
second particle center. While one cannot expect the calculations for 
thermal systems to hold also for static granular packings, the structure 
factors show generic features and trends like their principal peaks 
increasing with density that are also observed for packings.

Following Ornstein-Zernike, the structure factors for hard discs can be 
expressed as $S(q,\eta) = 1/(1 - 4\eta/\pi c(q,\eta))$, where $c(q,\eta)$ 
is the Fourier transform of a direct correlation function $c(r,\eta)$, $r$ 
the particle distance scaled by the diameter and $\eta$ the area fraction. 
M. Baus and J.-L. Colot \cite{Baus1986} have proposed an analytical 
formulation for the direct correlation function which yields typically 
good agreement with experiments especially in two dimensions,
\begin{equation}
	c(r,\eta) = \Theta(1-r)c(r=0,\eta)(1-a^{2}\eta(1+\omega(r/a)),
\label{eq:cr}
\end{equation} 
where $\Theta$ is the Heavyside function, $\omega(x)= (2/\pi)[cos^{-1}x - 
x(1-x^{2})^{1/2}]$ and $a$ is a scaling factor derived from thermodynamic 
arguments. We adjusted the area fraction to match the theoretical 
prediction for $S(q, \eta)$ with the experimental data. 


\begin{thebibliography}{10}
\providecommand{\url}[1]{\texttt{#1}}
\providecommand{\urlprefix}{URL }
\providecommand{\eprint}[2][]{\url{#2}}

\bibitem{Majmudar2007}
T. S. Majmudar and M. Sperl and S. Luding and R. P. Behringer,
\newblock Phys. Rev. Lett. \textbf{98,} 058001 (2007).

\bibitem{Jerkins2008}
M. Jerkins, M. Schr\"oter, H. L. Swinney, T. J. Senden, M. Saadatfar, and T. 
Aste,
\newblock Phys. Rev. Lett. \textbf{101,} 018301 (2008).

\bibitem{Donev2005}
A. Donev, F. H. Stillinger, and S. Torquato,
\newblock Phys. Rev. Lett. \textbf{95,} 090604 (2005).

\bibitem{Pusey1986}
P. N. Pusey and W. van Megen, 
\newblock Nature \textbf{320,} 340 (1986). 

\bibitem{Hubers2012}
E.~B\"undermann, H.-W. H\"ubers, and M.~Kimmitt,
\newblock \emph{Terahertz Techniques}.
\newblock (Springer, 2012).

\bibitem{Rhodes1991}
M.~J. Rhodes (Ed.),
\newblock \emph{Principles of Powder Technology}.
\newblock (John Wiley \& Sons, 1991).

\bibitem{Duran1999}
J.~Duran,
\newblock \emph{Sands, Powders, and Grains: An Introduction to the Physics of
Granular Materials}.
\newblock (Springer, 1999).

\bibitem{Campbell2003}
N.~A. Campbell (Ed.),
\newblock \emph{Biology: Exploring Life}.
\newblock (Pearson Prentice Hall, 2003).

\bibitem{Zurk2007}
L.~M. Zurk, B.~Orlowski, D.~P. Winebrenner, E.~I. Thorsos, M.~Leahy-Hoppa, 
and M.~R. Hayden, 
\newblock J. Opt. Soc. Am. B \textbf{24,} 2238 (2007).

\bibitem{Bandyopadhyay2007}
A.~Bandyopadhyay, A.~Sengupta, R.~B. Barat, D.~E. Gary, J.~F. Federici,
M.~Chen, and D.~B. Tanner,
\newblock J Infrared Milli Terahz Waves \textbf{28,} 969 (2007).

\bibitem{Kaushik2012a}
M.~Kaushik, B.~W.-H. Ng, B.~M. Fischer, and D.~Abbott,
\newblock Appl. Phys. Lett. \textbf{100,} 011107 (2010).

\bibitem{Kaushik2012b}
M.~Kaushik, B.~W.-H. Ng, B.~M. Fischer, and D.~Abbott,
\newblock Appl. Phys. Lett. \textbf{100,} 241110 (2012).

\bibitem{Kohler2002}
R.~K\"{o}hler, A.~Tredicucci, F.~Beltram, H.~E. Beere, E.~H. Linfield, A.~G.
  Davies, D.~A. Ritchie, R.~C. Iotti, and F.~Rossi,
\newblock Nature \textbf{417,} 156 (2002).

\bibitem{Hubers2010}
H.~Richter, M.~Greiner-B\"ar, S.~G.~Pavlov, A.~D.~Semenov, M.~Wienold, 
L.~Schrottke, M.~Giehler, R.~Hey, H.~T.~Grahn, and H.-W. H\"ubers,
\newblock Opt. Express \textbf{18,} 10177 (2010).

\bibitem{Cunningham2011}
P.~D. Cunningham, N.~N. Valdes, F.~A. Vallejo, L.~M. Hayden, B.~Polishak, 
X.-H. Zhou, J.~Luo, A.~K.-Y. Jen, J.~C. Williams, and R.~J. Twieg,
\newblock J. Appl. Phys. \textbf{109,} 043505 (2011).

\bibitem{Egelhaaf2006}
S.~U. Egelhaaf,
\newblock \emph{Solution Scattering}.
\newblock In \emph{{Soft condensed matter physics in molecular and cell 
biology}}, ed. by W.~C.~K. Poon and D. Andelman, p. 38. (Taylor \& 
Francis, 2006).

\bibitem{Hulst1981}
H.~C. van~de Hulst,
\newblock \emph{Light Scattering by Small Particles}.
\newblock (Dover Publications, 1981).

\bibitem{Matzler2002}
C.~M\"atzler,
\newblock \emph{{MATLAB Functions for Mie Scattering and Absorption}}.
\newblock IAP Research Report, Institut f\"ur angewandte Physik, 
Universit\"at Bern \textbf{08,} (2002).

\bibitem{BH1983}
C.~F. Bohren and D.~R. Huffman,
\newblock \emph{Absorption and Scattering of Light by Small Particles}.
\newblock (Wiley-VCH, 1983).

\bibitem{Wiscombe1980}
W.~J. Wiscombe,
\newblock Appl. Opt. \textbf{19,} 1505 (1980).

\bibitem{Brown1996}
W.~Brown (Ed.),
\newblock \emph{Light scattering : principles and development}.
\newblock (Clarendon Press, 1996).

\bibitem{Baus1986}
M.~Baus and J.~L. Colot,
\newblock J. Phys. C \textbf{19,} L643 
(1986).

\end{thebibliography}
\end{document}